\documentclass[twocolumn]{aastex63}

\usepackage{epsf}
\usepackage{amsmath}                
\usepackage{amsfonts}               
\usepackage{amssymb}                
\usepackage{epsfig}                 
\usepackage{float}
\usepackage{color}
\usepackage{tabularx}
\usepackage{comment}
\usepackage{makecell}
\usepackage{lineno}

\def \s{~\rm{s}}
\def \km{~\rm{km}}

\def \K{~\rm{K}}

\def \g{~\rm{g}}
\def \G{~\rm{G}}

\begin{document}


\title{Circumbinary disks in post common envelope binary systems with compact objects}


\author[0009-0004-5313-4341]{Lotem Unger}
\affiliation{Department of Physics, Technion - Israel Institute of Technology, Haifa, 3200003, Israel; lotem.unger@campus.technion.ac.il;  soker@physics.technion.ac.il}

\author[0000-0002-2215-1841]{Aldana Grichener}
\affiliation{Steward Observatory, University of Arizona, 933 North Cherry Avenue, Tucson, AZ 85721, USA; agrichener@arizona.edu}

\author[0000-0003-0375-8987]{Noam Soker}
\affiliation{Department of Physics, Technion - Israel Institute of Technology, Haifa, 3200003, Israel; lotem.unger@campus.technion.ac.il;  soker@physics.technion.ac.il}

\begin{abstract}
We conduct a population synthesis study using the binary population synthesis code \textsc{compas} to explore the formation of circumbinary disks (CBDs) following the common envelope evolution (CEE) phase of a giant star and a neutron star (NS) or black hole (BH). We focus on massive binary systems that evolve into double compact object (DCO) binaries after the exposed core of the giant collapses to form a second NS or BH. A CBD around the binary system of the giant's core and the compact object alters the orbital evolution of the binary. We parameterize the conditions for CBD formation in post-CEE binaries and present characteristics of DCO progenitors that are likely or unlikely to form CBDs. We find that CBD formation is most common in BH-BH binaries and NS-NS binaries that are expected to merge within Hubble time. Furthermore, we find that the interaction of the CBD with the core - NS/BH system at the termination of the CEE reduces the expected rate of DCO mergers, regardless of whether these binaries tighten or expand due to this interaction. If the binary system loses angular momentum to the CBD, it may produce a luminous transient due to a merger between the NS/BH and the core of the giant rather than gravitational wave sources. Thus, accounting for post-CEE CBD formation and its interaction with the binary system in population synthesis studies is significant for obtaining reliable predictions of the gravitational wave event rates expected by current detectors. 
\end{abstract}

\keywords{binaries: general -- stars: neutron stars -- stars: black holes  -- stars: massive   -- stars:supernovae}

\section{INTRODUCTION}
\label{sec:intro}

The common envelope evolution (CEE) phase involves a wide range of processes on various timescales and length-scales, many of which are very difficult to simulate even with the most sophisticated existing hydrodynamical numerical codes (for a review see, e.g.,  \citealt{RopkeDeMarco2023}). Three-dimensional hydrodynamical simulations of CEE over four decades (e.g., \citealt{DeKool1987, LivioSoker1988, Termanetal1994, RasioLivio1996, Sandquistetal1998, RickerTaam2008, Passyetal2012, Nandezetal2014, IvanovaNandez2016, Ohlmannetal2016, Staffetal2016, Iaconietal2017, PrustChang2019, Sandetal2020, GlanzPerets2021a, GlanzPerets2021b, Lauetal2022, Hilleletal2023, Schreieretal2023, BermudezBustamanteetal2024, Chamandyetal2024, Gagnieretal2024, GonzalezBolivaretal2024,  Landrietal2024, RosselliCalderon2024, Vetteretal2024}) and of the grazing envelope evolution (e.g., \citealt{Shiberetal2017, Shiberetal2019, ShiberIaconi2024}), have had difficulties including all the CEE sub-phases together and all physical processes required to fully simulate the CEE phase. 

One process the simulations have difficulties following is the possible formation of a circumbinary disk (CBD) around the surviving binary system. If the dynamical plunge-in of the accretor is stalled, the remnant binary could interact with the flattened envelope outside the binary orbit over a much longer timescale, resulting in the formation of a CBD (e.g., \citealt{KashiSoker2011}). The re-distribution of angular momentum and energy in the envelope gas leads to a torus-like structure around the core of the primary and the secondary stars. Current three-dimensional hydrodynamical simulations of the CEE face challenges in capturing this process in detail. While many of them find that after the ejection of the envelope its gas remains nearby forming a flattened envelope (e.g., \citealt{Sandquistetal1998}) or a collimated bipolar structure along the symmetry axis (e.g., \citealt{Chamandyetal2018, Chamandyetal2020, Zouetal2020, Zouetal2022, Ondratscheketal2022}) as previously predicted by an analytical calculation \citep{Soker1992funnel}, showing the potential of CBD formation, simulations resolving the torus-like structure (e.g., \citealt{Vetteretal2024}) are scarce.    

The interaction of the post-CEE CBD with the binary system was suggested to influence the subsequent binary evolution  (e.g., \citealt{KashiSoker2011, ChenPodsiadlowski2017, Munozetal2019, DOrazioDuffell2021, TunaMetzger2023, GagnierPejcha2023, Weietal2024, Siweketal2023, Vallietal2024, Vetteretal2024}). Hydrodynamical simulations that aim to follow the evolution of CBDs and their long-term interaction with the binary system they encompass encounter multiple numerical challenges. For instance, properly resolving the cavity that contains the binary is crucial for estimating the torques that dictate the orbital evolution (e.g., \citealt{GagnierPejcha2023}). Since the core of the giant star and its companion are usually represented as point masses with artificial gravitational potentials, the gravitational torques are inherently inaccurate, leading to imprecise orbital separations. Another layer of complexity in simulating the CBD-binary interaction is posed by correctly modeling magnetic fields in the post-dynamical in-spiral phase of CEE (e.g., \citealt{GagnierPejcha2024}). Although magnetic fields are thought to have negligible impact on binary orbital separations, their amplification could lead to outflows that alter the morphology and densities around the disk, influencing the orbital evolution. 
 
In post-common envelope systems of massive stars, the interaction of the CBD with the binary could lead to detectable transients. If the binary system contains a neutron star (NS) or a black hole (BH) and the stripped core of a giant, then mass accretion by the compact object from the CBD could result in the launching of jets that power a luminous event.  \cite{TunaMetzger2023} suggest that jet-powered radio emission from such a system might be compatible with expectations from fast blue optical transients. In particular, long-lived disks could result in delayed explosions with less hydrogen as in the case of AT2018cow-like transients or type Ibn/Icn supernovae (e.g., \citealt{Soker2022}; \citealt{Metzger2022}; \citealt{TunaMetzger2023}; \citealt{Weietal2024}). Furthermore, dust formation by the jets launched from the CBD could result in detectable IR radiation (see \citealt{Grichener2024} for a discussion on potential observational signatures of such systems). In the case of much wider post-asymptotic giant branch binaries, having orbital periods of $\simeq 100-1000$~days and masses of $M_1\simeq 1 M_\odot$, mass accretion and jets launching from a CBD of several AU could shape a bipolar nebula (e.g., \citealt{Bollenetal2022, Verhammeetal2024}). These CBDs, however, are expected to be post-grazing envelope evolution disks rather than CEE (e.g., \citealt{Soker2020G}).

One approach to allow a wide exploration of CEE and post-CEE outcomes is parameterizing processes. This is particularly critical in performing population synthesis studies. We aim to incorporate the possibility of forming a post-CEE CBD into population synthesis; here, we re-analyze the data from \cite{Grichener2023} (publicly available at Zenodo: \href{https://zenodo.org/records/11237180}{doi:10.5281}) generated by the rapid population synthesis code \textsc{compas} \citep{TeamCOMPAS2022}, as described in Section \ref{sec:NumericalScheme}. Towards our goal, in Section \ref{sec:CBDFormation}, we build a toy model to parameterize the conditions to form post-CEE CBDs. In Section \ref{sec:ResultsCBDs} we employ our toy model to the population synthesis data of massive binary systems that end with either NS - NS, BH - BH, or NS - BH binary systems. In Section \ref{sec:DCOsAndLFBOTs}, we discuss the potential implications of our results for the formation and merger rates of double compact object (DCO) binaries. We summarize and discuss the effect of extending the parameter space on our results in Section \ref{sec:Summary}. 

\section{Methods}
\label{sec:NumericalScheme}
We use rapid population synthesis to follow the evolution of massive binaries and find the properties of systems where a circumbinary disk would form in the post-CEE phase of a compact object (NS or BH) and a giant star. We analyze datasets from \cite{Grichener2023}\footnote{Publicly available at Zenodo: \href{https://zenodo.org/records/11237180}{doi:10.5281}.} generated by version 02.31.06  of the population synthesis {code} \textsc{compas} (Compact Object Mergers: Population Astrophysics and Statistics; \citealt{Stevensonetal2017}; \citealt{VignaGomezetal2018}; \citealt{TeamCOMPAS2022}). 

Given initial distributions and parametric assumptions, \textsc{compas} evolves isolated binary systems using the analytical fits from \cite{Hurley2000} and \cite{Hurleyetal2002}. The mass distribution of the primary (initially more massive) star is determined by the Kroupa initial mass function in the form $dN/dM_{\rm 1,ZAMS}\propto M_{\rm 1,ZAMS}^{-2.3}$ \citep{Kroupa2001}. The range of the masses $5 \leq M_{\rm 1,ZAMS}/M_{\odot} \leq 100$ was chosen to account for primaries that could result in a supernova explosion and the formation of a compact object.\footnote{The stars with the lowest masses in our distribution can gain mass through mass transfer and become massive enough to explode as supernovae.} The mass of the secondary (initially less massive) star was determined by a flat distribution for a mass ratio between $0.1 \leq q_{\text{ZAMS}} \equiv M_{2,\text{ZAMS}} \slash M_{1,\text{ZAMS}} \leq 1$ and the semi-major axis was determined by a flat-in-log distribution in the range $0.1 \leq a_{\rm ZAMS} \rm /AU \leq 1000$ \citep{Sana2012}. For the NS natal-kicks induced by supernova explosions we take a bimodal velocity distribution with a higher mode of $\sigma_{\rm high}=265 \km \s^{-1}$ for regular core collapse supernovae and a lower mode of $\sigma_{\rm high}=30 \km \s^{-1}$ corresponding to electron capture supernovae and ultra stripped supernovae, while for the BHs in our sample we reduce this natal-kick distribution by a fallback factor. We focus on sub-populations of $10^{7}$ binaries with solar metallicity ($Z_{\rm \odot}=0.0142$; \citealt{Asplund2009}) and common envelope efficiency parameter $\alpha_{\rm CE}=1$ (equation \ref{eq:EnergyFormalism}) as in the 'fiducial' model of \cite{VignaGomezetal2018}. 

In some of the systems sampled in {\cite{Grichener2023}} there were episodes of dynamically unstable mass transfer that led to CEE. 
\textsc{compas} determines the stability of mass transfer by comparing the donor's response to mass loss with that of the Roche-lobe radius: if the donor's radius response, $d \ln R_{d, MT} \slash d \ln M_{d, MT}$, where $ R_{d, MT}$ and $ M_{d, MT}$ are the radius and mass of the donor at the onset of mass transfer,  is greater than or equal to that of the Roche lobe, $d \ln R_{\mathrm{RL}} \slash d \ln M_{d,MT}$, where $R_{\mathrm{RL}}$ is the Roche radius, the mass transfer is stable. Otherwise, it results in a CEE event (e.g., \citealt{PaczynskiSienkiewicz1972}). To find the value of the semi-major axis immediately after the CEE phase, \textsc{compas} uses the energy formalism \footnote{Other methods use angular momentum parametrization to estimate the final orbital separation (e.g., \citealt{NelemansTout2005, Toonenetal2012, DiStefano2023}). However, the angular momentum prescriptions could be problematic  (e.g., \citealt{Webbink2008, CohenSoker2023b}).}. In this formalism, we compare the binding energy of the envelope $E_{\rm bind}$ to the orbital energy before and after the CEE phase (e.g., \citealt{Van1976, IbenTutukov1984, LivioSoker1988, Ivanovaetal2013, IaconiDeMarco2019}) 
\begin{equation}
\begin{split}
 E_{\rm bind} = 
 \frac{\alpha_{\rm CE}GM_{\rm 1,i}M_{\rm 2,i}}{2a_{\rm i}}  - \frac{ \alpha_{\rm CE}G M_{\rm 1,f} M_{\rm 2,f}}{2a_{\rm f}}  ,
\label{eq:EnergyFormalism}
\end{split}
\end{equation}
where $\alpha_{\rm CE}$ is the common envelope efficiency parameter that determines which fraction of the orbital energy will go to unbinding the envelope, $M_{\rm 1,i}$ and $M_{\rm 2,i}$ are the masses of the primary and secondary stars at the onset of CEE, respectively, $M_{\rm 1,f}$ and $M_{\rm 2,f}$ are the masses of the primary and secondary stars after CEE, respectively, $a_{\rm i}$ is the semi-major axis of the binary systems before the CEE phase and $a_{\rm f}$ is the semi-major axis of the binary systems at the end of CEE. The binding energy is computed following \cite{deKool1990} 
\begin{equation}
E_{\rm bind} = - \frac{GM_{\rm i} M_{\rm env,i}}{\lambda R_{\rm i}},
\label{eq:bindingEnergy}
\end{equation}
where $M_{\rm i}$, $M_{\rm env,i}$ and $R_{\rm i}$ are the mass, envelope mass and radius of the donor, respectively, and $\lambda$ is a structure parameter calculated by the prescription in \cite{XuLi2010a} and \cite{XuLi2010b}. The structure parameter includes contributions of the full internal energy, which consists of the thermal energy, radiation energy, ionization energy and dissociation energy of molecular hydrogen. For more information about the population synthesis model assumptions and initial distributions see Section 2 in \cite{Grichener2023}.

We focus on sub-populations in which a compact object enters a CEE phase with a giant star, where the formation of a CBD after CEE could potentially lead to shrinkage in the orbital separation and a post-CEE merger (e.g., \citealt{TunaMetzger2023}; \citealt{Weietal2024}), accounting for various luminous transients (e.g., \citealt{Grichener2024}). Without taking the formation and interaction of the binary with the CBD into consideration, most of these systems end as DCO binaries, implying a potential reduction in their formation and merger rates (see Section \ref{sec:DCOsAndLFBOTs}). \textsc{compas} allows us to examine which parameters and stellar quantities can affect the process of creating a CBD. To find the properties of the systems that will form a post-common envelope CBD, we compare the specific angular momentum of the envelope at the end of the Roche Lobe Overflow (RLOF) phase to the Keplerian specific angular momentum around the binary system, as we further elaborate in Section \ref{sec:CBDFormation}.

\section{Parametrization of post-common envelope circumbinary disk formation}
\label{sec:CBDFormation}
\subsection{The toy model}
\label{subsec:ToyModel}

We build a toy model to parameterize the conditions for forming a post-CEE CBD and incorporate it in population synthesis studies. We estimate the fraction of the angular momentum at the onset of RLOF that the leftover envelope retains and equate it to the minimum angular momentum that the leftover requires to form a CBD. We then implement this parametrization in the data of \cite{Grichener2023} and present our results in Sections \ref{sec:ResultsCBDs} and \ref{sec:DCOsAndLFBOTs}. 

We assume that at the entrance to RLOF the secondary star, an NS or a BH in our study, manages to spin up the giant's envelope to synchronization (e.g., \citealt{VignaGomezetal2020}). We take the envelope to have a solid-body rotation with the synchronization angular velocity $\omega = \sqrt{G(M_{\rm *,RL} + M_{\rm CO,RL})/a_{\rm RL}^{3}}$, which corresponds to the Keplerian velocity of the compact object at the Roche-radius, where $M_{\rm *,RL}$ and $M_{\rm CO,RL}$ are the mass of the giant and the mass of the compact object at this stage, respectively, and $a_{\rm RL}$ is the orbital separation. For the moment of inertia of the envelope at this stage, we take $I_{\rm env,RL} = k_{\rm e} M_{\rm env,RL} R^2_{\rm \ast, RL}$, where $M_{\rm env,RL}$ and $R_{*,RL}$ are the envelope mass and radius of the giant at the onset of RLOF, and $k_{\rm e} \simeq 0.2$ for extended envelopes of giants (e.g., \citealt{Soker2004}). The total angular momentum of the binary at the onset of RLOF is 
\begin{equation}
\begin{split}
J_{\rm total,RL} = J_{\rm orb,RL} + J_{\rm rot,RL} = \\ \mu_{\rm RL}\sqrt{G(M_{\rm *,RL}+M_{\rm CO,RL})a_{\rm RL}(1-e^{2})}+\omega I_{\rm env,RL},
\label{eq:j_total}
\end{split}
\end{equation}
where $j_{\rm orb,RL}$ is the orbital angular momentum, $j_{\rm rot,RL}$ is the rotational angular momentum of the envelope and $\mu\equiv M_{\rm *,RL} M_{\rm CO,RL} / (M_{\rm *,RL} + M_{\rm CO,RL})$ at this stage, and $e$ is the eccentricity of the orbit\footnote{In these systems we take $e$ prior to circularization at the onset of RLOF assumed in \textsc{compas}, which is often considerable due to the first SN's natal kick.} . We exclude the rotational angular momentum of the core and of the compact object, which are negligible compared to the rotational angular momentum of the envelope due to their extremely small radii.

Let us denote by $f_{\rm J}$ the fraction of angular momentum retained by the binary system after the CEE phase, and by $f_{\rm M}$ the fraction of mass that remains bound. We take the post-CEE specific angular momentum available to form a CBD to be a fraction $\beta \equiv f_{\rm J}/f_{\rm M} $ of the total specific angular 
momentum of the binary at the onset of RLOF (similarly to \citealt{IzzardJermyn2023})
\begin{equation}
\begin{split}
j_{\rm CBD} =  \frac{f_{\rm J} J_{\rm total,RL}}{f_{\rm M} M_{\rm env,RL}} = \beta \frac{J_{\rm total,RL}}{M_{\rm env,RL}} ,
\label{eq:beta_updated}
\end{split}
\end{equation}

Post-CEE CBDs typically form with inner radii $R_{\rm CBD,in}$ approximately 2–3 times larger than the orbital separation of the binary at the end of the CEE phase (e.g., \citealt{KashiSoker2011}; \citealt{TunaMetzger2023}). The absolute minimal specific angular momentum required to form a thin ring that would viscously spread into a centrifugally-supported CBD is set by $R_{\rm CBD,in}$ . However, unlike the classic \cite{LyndenBellPringle1974} disk description, which is less likely to form from CEE dynamics, we are interested in a disk which already extends from the inner stable orbit to a much larger radius, and thus has significantly larger angular momentum than a thin ring of the same mass at the inner stable orbit. Therefore, we must allow for additional angular momentum to support the gas that is further out. This more restrictive approach is motivated by the structure found in the hydrodynamical simulation of \cite{Vetteretal2024}, and is conceptually similar to assuming the circularization of the CBD gas at a larger effective radius. Our criterion for CBD formation is then
\begin{equation}
j_{\rm CBD} \gtrsim \zeta \sqrt{G(M_{\rm core}+M_{\rm CO,f}) R_{\rm CBD,in}},
\label{eq:jenvf}
\end{equation}
where $M_{\rm core}$ is the mass of the giant's core that becomes stripped after CEE, $M_{\rm CO,f}$ is the mass of the compact object companion at this stage, $R_{\rm CBD,in}=2.5a_{\rm f}$ and $\zeta$ is a dimensionless parameter that accounts for the mass distribution of the disk. For a disk whose surface density is a simple power law $\Sigma (r) = \Sigma_{\rm 0} (r/ r_{\rm 0})^{-m}$, where $\Sigma_{\rm 0}$ and $r_{\rm 0}$ are normalization constants, calculating the mean specific angular momentum gives $\zeta = \frac{m - 2}{m - 2.5}$ (see Appendix for the derivation). The hydrodynamical simulations performed by \cite{Vetteretal2024} yield a more complicated structure. In particular, their disk has a clear outer radius, beyond which the gas is flowing radially outward rather than around the binary. We take a fiducial value of $\zeta = 4$, which corresponds to $m \simeq 2.75$ and is roughly consistent with the density profile obtained by \cite{Vetteretal2024}.Setting $\zeta=4$ can equivalently be interpreted as assuming that the gas circularizes into a CBD with an inner radius $R_{\rm CBD,in}$ that is 32 to 48 times  the post CEE orbital separation. Such values are reasonable in systems with a large leftover envelope. Smaller values of $\beta$ will result in more DCOs that form post CEE CBDs. In addition, we assume that the mass of the disk is much smaller than the total mass of the binary system, and hence we can neglect its self gravity. 

\subsection{The role of the parameters $\beta$ and $\zeta$}
\label{subsec:Values}
 
To demonstrate the role of $\beta$ in determining the number of post-CEE systems that form a CBD, we re-analyze the population synthesis data from \cite{Grichener2023} and examine three final populations that go through the CEE of a giant star with a compact object, and end as NS-NS, BH-BH, and NS-BH binaries. 

\begin{figure}
\begin{center}
\hspace*{-0.3cm}
\includegraphics[width=0.49\textwidth]{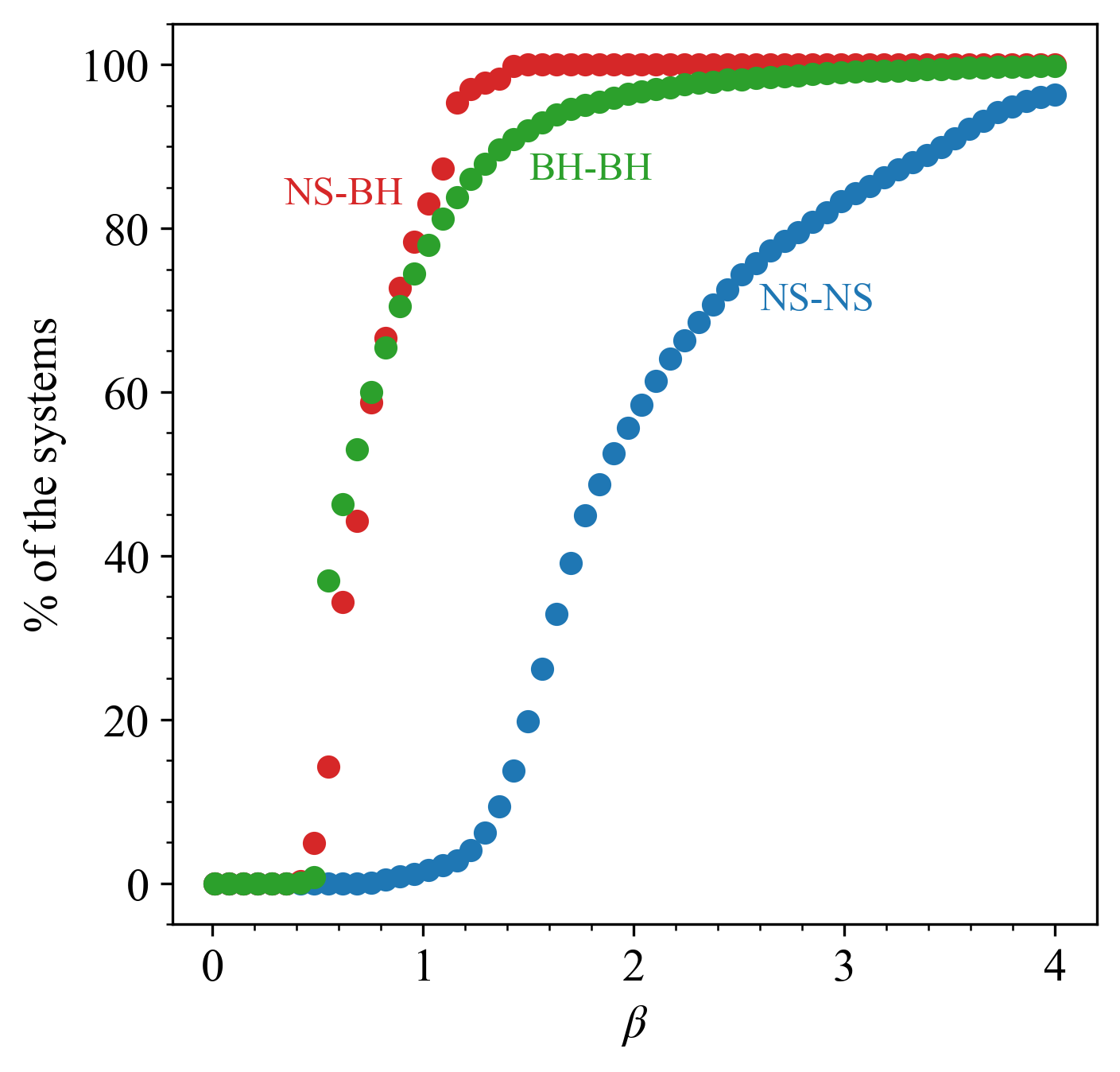}
\caption{The percentage of NS-NS (blue dots), BH-BH  (green dots), and NS-BH (red dots) systems that can form a CBD, from all systems that go through NS/BH-giant CEE phase and end as NS-NS systems, BH-BH systems, and NS-BH systems, respectively, taking a more restrictive condition for CBD formation ($\zeta = 4$). Assuming the minimum specific angular momentum needed to form a thin ring that can subsequently viscously spread into a CBD (i.e., $\zeta = 1$), the corresponding $\beta$-axis should be rescaled downwards by a factor of four. } 
\label{fig:allbeta}
\end{center}
\end{figure}

In Figure \ref{fig:allbeta}, we present the percentage of NS-NS systems (blue dots), BH-BH systems (green dots), and NS-BH systems (red dots) that form a post-CEE CBD as a function of the parameter $\beta$. As expected, higher values of $\beta$, which represent systems that retain more of their initial angular momentum, correspond to a greater percentage of systems forming a post-CEE CBD. For BH-BH binaries and NS-BH binaries we note a sharp transition in $\beta$ between systems that are likely to form CBDs and those where such disks are not expected, while for NS-NS binaries the dependence is weaker. While recent papers (e.g., \citealt{IzzardJermyn2023}) show a preference towards high values of $\beta$, due to the large uncertainties involved we show the results of a case where half of the systems will form CBDs. When varying the value of $\beta$, we find that qualitative trends in our results would change only if most systems formed CBDs. The graph shows that a CBD is formed in about half of the systems for $\beta = 0.656$ in BH-BH binaries, $\beta = 1.859$ in NS-NS binaries, and $\beta = 0.713$ in NS-BH binaries. 

We note from equations \ref{eq:beta_updated} and \ref{eq:jenvf} that the parameter $\zeta$, which accounts for the mass distribution of the CBD, scales linearly with $\beta$. This implies that adopting a different mass distribution (i.e., a different $\zeta$) rescales the value of $\beta$ required to produce a fixed number of CBDs in each sub-population by the same multiplicative factor. Assuming a thin ring that viscously spreads to a CBD, corresponding to the absolute minimum mean specific angular momentum required for CBD formation, for instance, is represented by $\zeta=1$, and yields $\beta$ values that are a factor of four smaller than those quoted above for $\zeta=4$.

\section{Properties of post-common envelope circumbinary disk progenitors}
\label{sec:ResultsCBDs}

Here, we explore the pre and post mass transfer characteristics of binary systems that result in DCOs and compare binaries that are likely to form CBDs after the termination of CEE with binaries that are not. 

\begin{figure*}
\begin{center}
\hspace*{-0.3cm}
\includegraphics[width=1.05\textwidth]{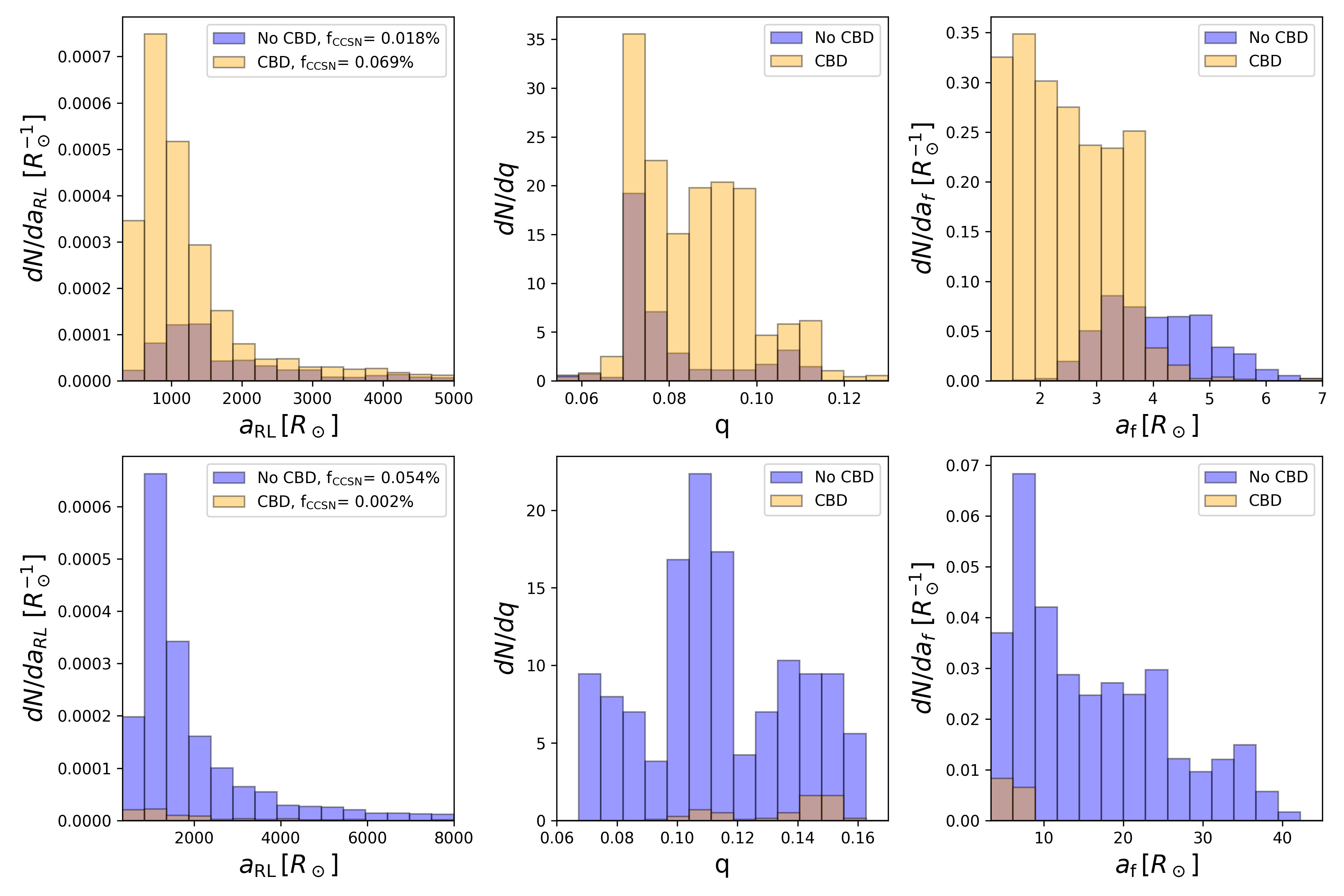}
\caption{  Properties of NS-NS binary systems for $\beta=1.859$ (see equation \ref{eq:beta_updated}) with (orange bins) and without (blue bins) post-CEE CBDs. The histograms show the distributions of the orbital semi-major axis $a_{\rm RL}$ (left panels), and the mass ratio $q \equiv M_{\mathrm{CO},RL}/ M_{\rm *,RL}$ (middle panels) at the onset of RLOF, and of the orbital separation after CEE $a_{\rm f}$. The vertical axis represents the number of systems per bin, normalized such that the total sum over all bins is equal to one. The upper row displays the distributions of systems that will merge within Hubble time and the bottom row presents distributions for systems that are not expected to merge. The factor $f_{\rm CCSN}$ is the fraction of the systems relative to all core-collapse supernovae. } 
\label{fig:Fig_NSNS}
\end{center}
\end{figure*}

\begin{figure*}
\begin{center}
\hspace*{-0.3cm}
\includegraphics[width=1.05\textwidth]
{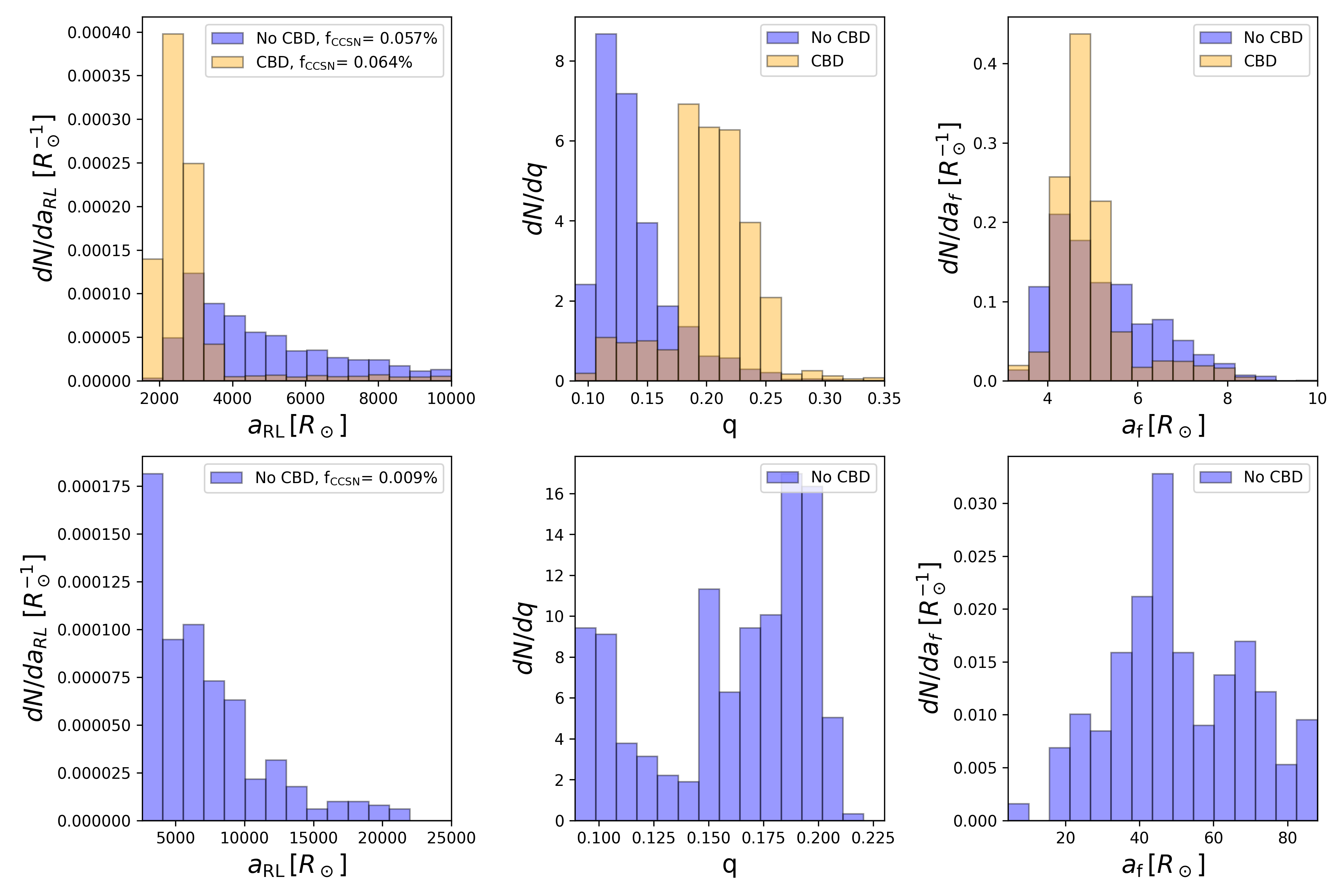} 
\caption{ 
Similar to Figure \ref{fig:Fig_NSNS} for BH-BH binary systems with $\beta$=0.656 .} 
\label{fig:Fig_BHBH}
\end{center}
\end{figure*}

\begin{figure*}
\begin{center}
\hspace*{-0.3cm}
\includegraphics[width=1.05\textwidth]
{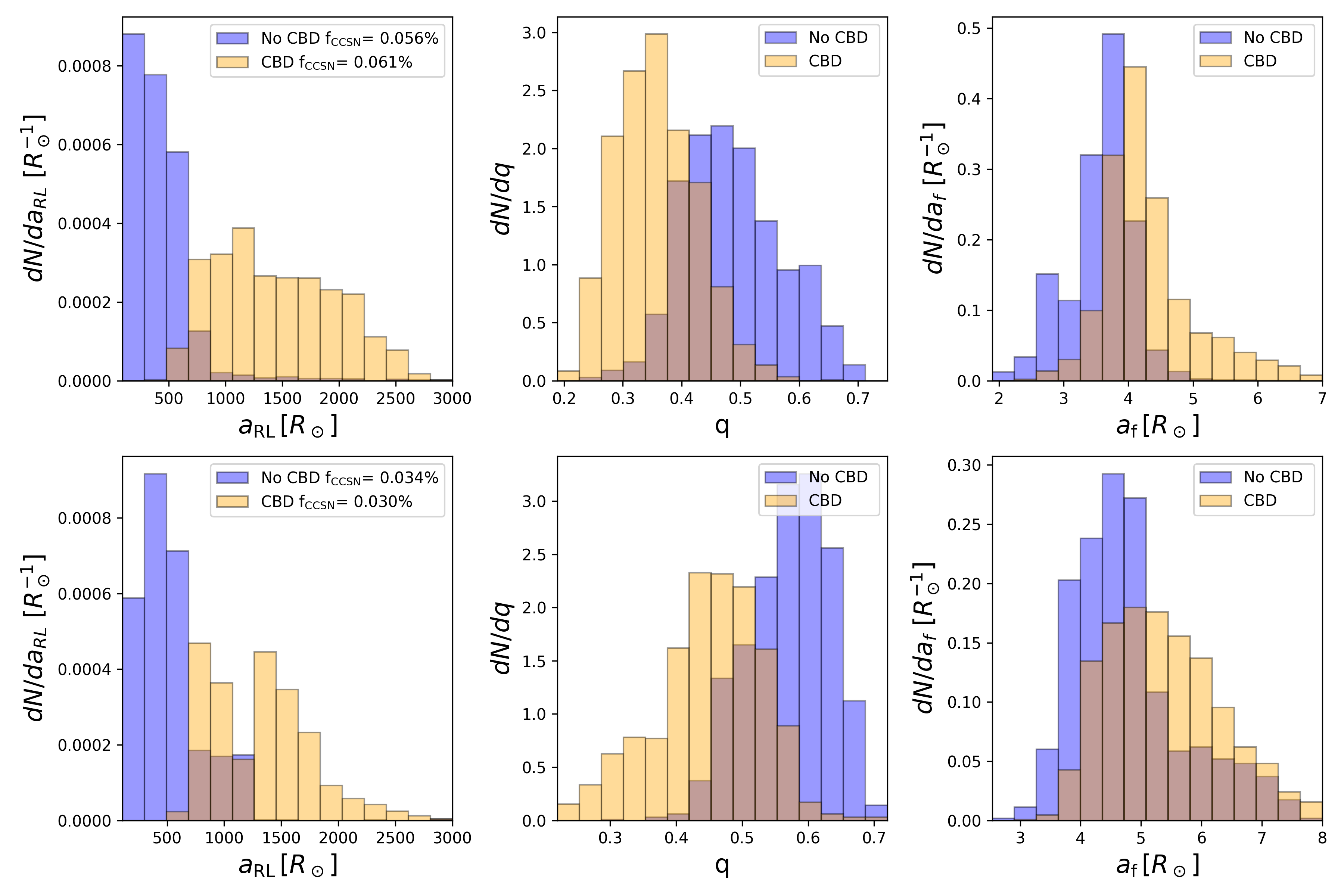} 
\vspace*{-0.1cm}
\caption{  
Similar to Figure \ref{fig:Fig_NSNS} for NS-BH binary systems with $\beta$=0.713 . }
\label{fig:Fig_NSBH}
\end{center}
\end{figure*}

In Figures \ref{fig:Fig_NSNS}-\ref{fig:Fig_NSBH}, we present the properties of NS-NS for $\beta=1.859$, BH-BH for $\beta=0.656$, and NS-BH for $\beta=0.713$ binaries, respectively. These values of $\beta$ indicate that about half of the systems have a post-CEE CBD (Figure \ref{fig:allbeta}). Orange bins represent the distributions for cases where the leftover flattened post-CEE envelope has enough angular momentum to form a CBD according to the toy model presented in Section \ref{subsec:ToyModel}. Blue bins are for binary systems that form no CBDs. The upper panels of each figure show DCO binaries that will merge within Hubble time by emitting gravitational waves. In contrast, in the lower panels, we show binaries where the compact objects are too far apart to merge. The factor $f_{\rm CCSN}$ displayed in the left insets of Figures \ref{fig:Fig_NSNS}-\ref{fig:Fig_NSBH} is the percentage of each sub-class of systems from all core-collapse supernovae in our sample, which we present for observational reasons. The interaction of the CBD with the binary system might lead to a luminous transient triggered by the merger of the NS or the BH with the exposed core (see section \ref{sec:DCOsAndLFBOTs} for a discussion).
This transient event is termed common envelope jets supernova (CEJSNe; e.g., \citealt{SokerGilkisiPTF14hls}). At an earlier stage, when the NS/BH accretes mass in the envelope and launches jets, the bright transient is called a CEJSN impostor (e.g., \citealt{Gilkisetal2019}). Initially, a CEJSN or a CEJSN impostor might be misclassified as a CCSN. Later peculiar properties, e.g., no radioactive products (for CEJSN impostors), longer activity, and larger total energy than typical CCSNe, make the differentiation.   

The lower panels of Figures \ref{fig:Fig_NSNS}-\ref{fig:Fig_NSBH} show that most NS-NS and BH-BH binaries that will not merge in Hubble time do not form post-CEE CBDs, while $\simeq 25\%$ of non-merging NS-BH binaries form CBDs after CEE of the BH and the NS progenitor. By definition, for a high value of $\beta$ all double compact objects will form CBDs, including non-merging NS-NS and BH-BH binaries. However, we find that it is less likely for a substantial fraction of non-merging DCOs to form CBDs, since it would require high values of $\beta$.

In the left and middle panels of Figures \ref{fig:Fig_NSNS}-\ref{fig:Fig_NSBH} we show the semi-major axis and mass ratio of the binary systems at the onset of RLOF, respectively. The progenitors of NS-NS binaries tend to have smaller initial orbital separations and mass ratios, in contrast to the case of BH-BH binaries and NS-BH binaries, which might be a source of difference in the results. However, despite the fact that we would intuitively think that these two separations would play a strong role in determining the angular momentum reservoir available for CBD formation, they do not separate systems that form post-CEE CBDs from those who do not. This is due to the non-trivial dependence of the angular momentum on the different masses involved in the problem. The right panels of figures \ref{fig:Fig_NSNS}-\ref{fig:Fig_NSBH} show that the final orbital separation of the binary, that determines that boundaries of the disk in case it forms (Section \ref{subsec:ToyModel}), is not a determining factor in CBD formation as well.

  
\section{Implications for double compact object binaries and transients}
\label{sec:DCOsAndLFBOTs}

When an NS or a BH is close enough to a giant star, they can engage in a CEE phase where the compact object unbinds the envelope gas as it spirals in towards the center of the giant star. In cases where the entire envelope is ejected, the stripped core and the compact object remain in a contact binary, where the core could explode as a stripped-envelope supernova (e.g., \citealt{Podsiadlowskietal1992}; \citealt{Laplaceetal2021SN}; \citealt{Vartanyanetal2021SN}; \citealt{Luetal2025}), forming another compact object. If the natal kick from the supernova is low enough, the binary system remains bound, resulting in a DCO binary. However, forming a post-CEE CBD encompassing the core and the compact object and its interaction with the binary system might affect the orbital evolution. If the binary system loses angular momentum to the CBD through gravitational torques, the orbital separation will shrink, altering the formation and merger rates of DCO binaries in several ways. 

\begin{figure*}
\begin{center}
\hspace*{-0.3cm}
\includegraphics[width=0.85\textwidth]
{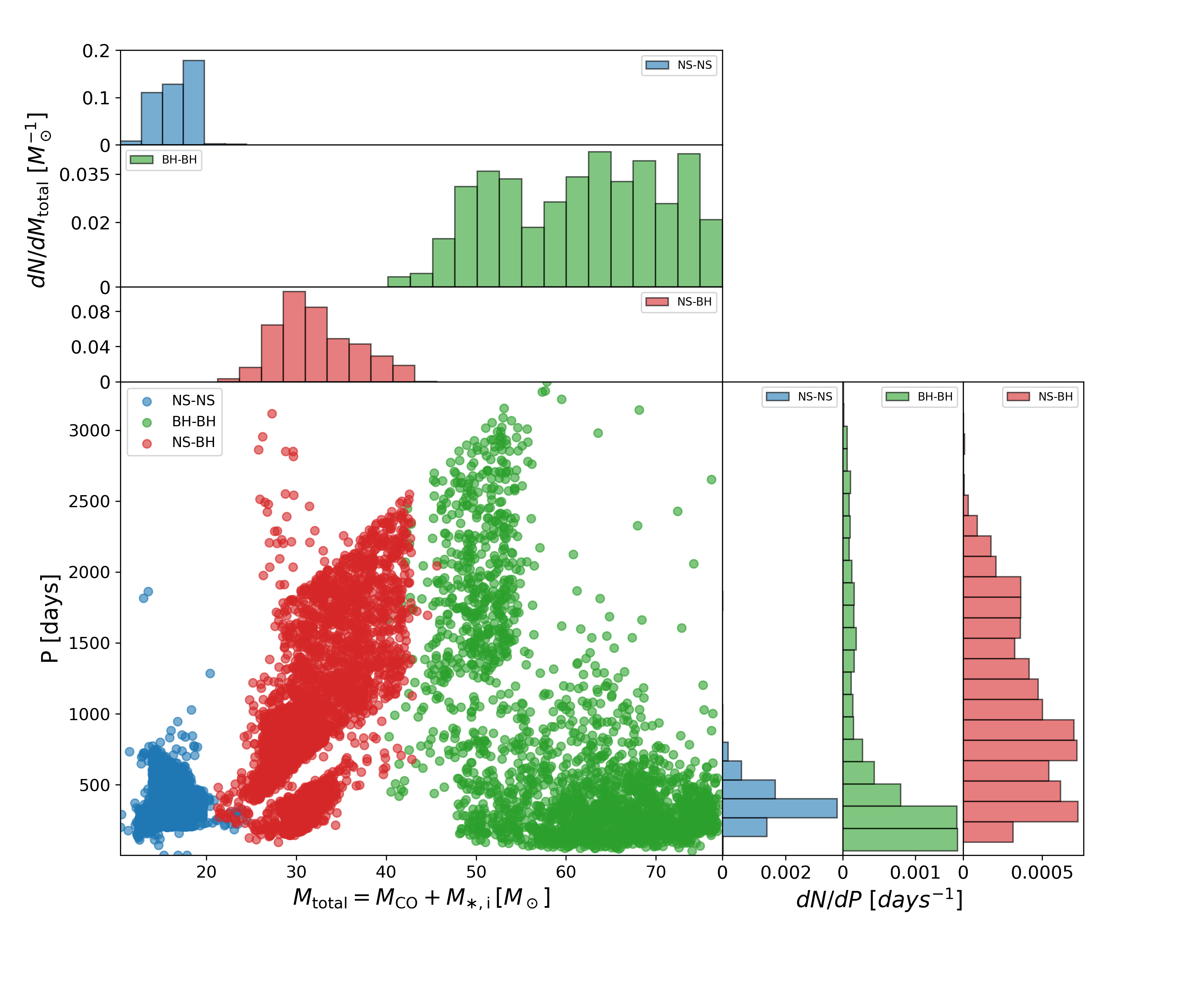} 
\vspace*{-1cm}
\caption{ 
Orbital period vs total mass of the binary systems at the onset of CEE for progenitors of NS-NS binaries (blue dots), BH-BH binaries (green dots), and NS-BH binaries (red dots) that form a post-CEE CBD. Above and to the right: histograms showing the distributions of both quantities for the different DCO binaries subclasses. We normalized the histograms such that in each sub-panel the sum over the bins is equal to one.
}.  
\label{fig:Fig_time_period}
\end{center}
\end{figure*}

In cases where the compact object and the core are relatively close at the end of CEE, the interaction with the CBD could lead to a merger between the core and the compact object (e,g., \citealt{Weietal2024}), giving rise to luminous transients, CEJSNe, appearing as, e.g., fast blue optical transients and long gamma-ray bursts (e.g.,  \citealt{Sokeretal2019}; \citealt{Soker2022}; \citealt{Metzger2022} \footnote{We note that the CBDs studied in \cite{Metzger2022} to account for fast blue optical transients are much wider than the CBDs we explore here.}; \citealt{Grichener2024}), reducing the number of systems that continue evolving towards DCO binaries. Figure \ref{fig:Fig_time_period} shows the predicted periods of binary systems that form post-CEE CBDs and hence have the potential to trigger these transients as a function of the total binary mass at the onset of CEE.  Since CEJSNe might be classified as peculiar CCSNe, at least at early times of the event, it is important to present the properties of their progenitors for future identifications in observations. Figure \ref{fig:Fig_time_period} further emphasizes the properties of the systems we study, which can be used to re-analyze peculiar CCSNe and reclassify them as CEJSN events.

If the stripped core and the compact object begin further apart, however, then a compact object-core merger is prevented and the system would still evolve towards a DCO binary. The reduction in the orbital separation could lead to DCO merger in Hubble time and gravitational wave emission in systems that would have been too far to merge without the interaction with the CBD (e.g., \citealt{Weietal2024}; \citealt{Siweketal2023} \citealt{Vetteretal2024}). In the case of a BH-BH binary system, the merger might result in detectable long gamma-ray burst \citep{Janiuketal2013}.

\begin{figure*}
    \begin{center}
        \includegraphics[width=1\textwidth]{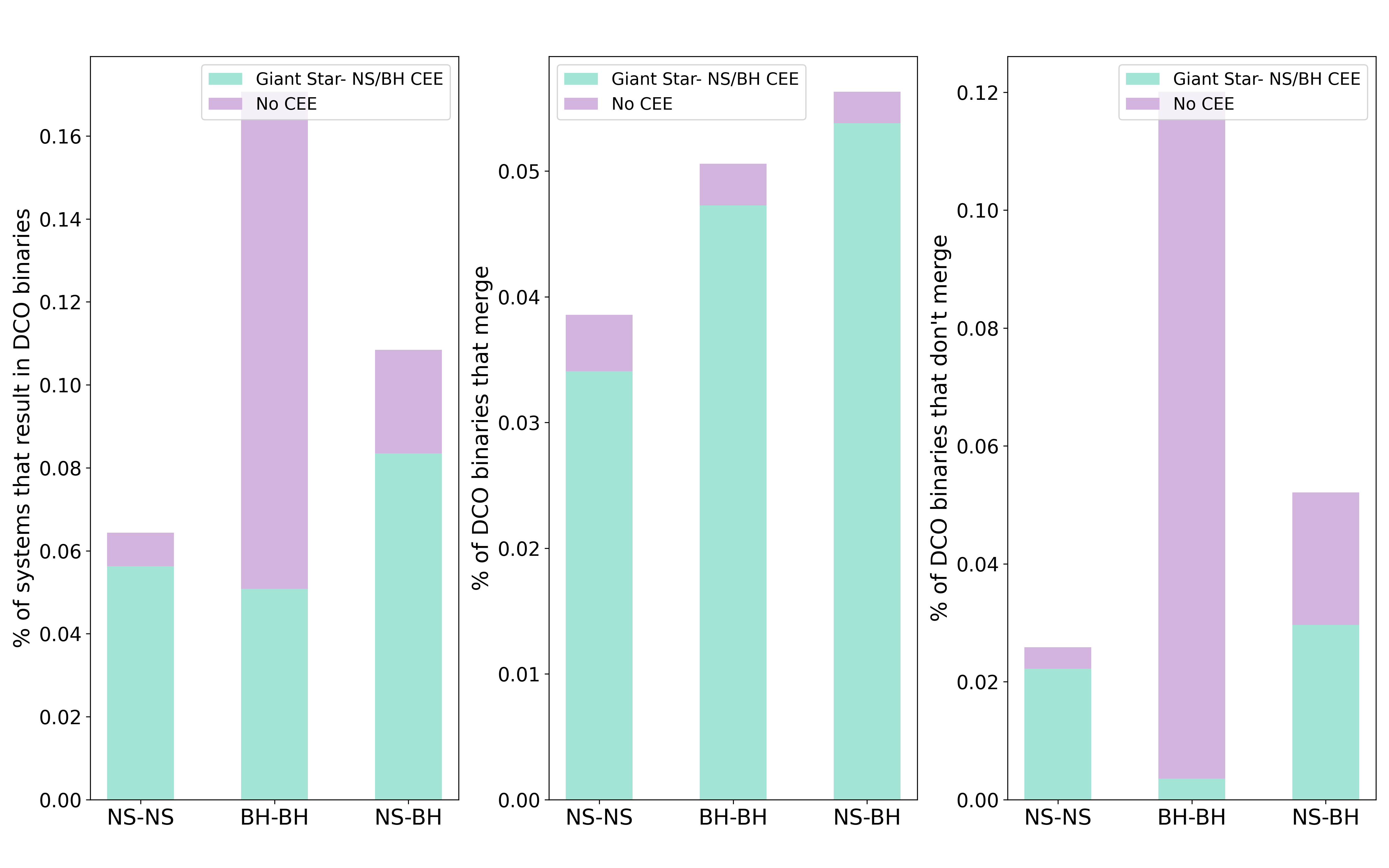}
        \caption{Percentage of binary systems that evolve to DCO binaries from the overall number systems in our binary stellar population, i.e., all primary stars in the range $5 \le M_{\rm 1,ZAMS} \le 100$. The left panel shows all DCO binaries that went through a CEE phase between the compact objects and the giant star throughout their evolution (green columns; See Figure 2 in \citealt{Grichener2023}), or that evolved via other channels (purple columns). The middle and right panels present the same information for DCOs that merge within Hubble time and for DCO binaries that are too wide to merge, respectively. }
        \label{fig:BinaryData}
    \end{center}
\end{figure*}

In the left panel of Figure \ref{fig:BinaryData}, we show the percentage of systems that result in NS-NS, BH-BH, and NS-BH binary systems in our stellar population, namely, all binary stars with primary initial mass in the range $5 \le M_{\rm 1,ZAMS} \le 100$ . The green columns represent DCO binaries that evolved through a CEE phase between the first compact object and the giant star. In contrast, the purple columns indicate systems formed via alternative evolutionary channels. We distinguish between systems in which the binary is expected to merge within Hubble time (middle panel) and in which the stars are not close enough to merge (right panel). While most NS-NS and NS-BH systems involve a giant-compact object CEE, this is not the case for BH-BH binaries, where most systems do not experience this evolutionary phase (as seen in the left panel). In the majority of DCOs that merge, however, CEE is essential to decrease the orbital separation (middle panel; similarly to other population synthesis studies like e.g.,  \citealt{Mapellietal2017}; \citealt{VignaGomezetal2018};  \citealt{Giacobboetal2018}; \citealt{Chruslinskaetal2018};  \citealt{Neijsseletal2019}; \citealt{Broekgaardenetal2021}; \citealt{MandelBroekgaarden2022} and references therein; see \citealt{GallegosGarciaetal2021} for a different conclusion drawn based on detailed stellar evolution simulations). 
   
Let us analyze each class of DCO binaries separately. The vast majority of NS-NS binary systems, whether they are expected to merge or not, are formed through an evolutionary pathway that includes CEE between an NS and a giant star (Figure \ref{fig:BinaryData}). Since the Keplerian specific angular momentum at the expected inner radius of the disk in post-CEE binary systems that would evolve to non-merging NS-NS binaries is high compared to the angular momentum of the flattened envelope, a CBD is not expected to form in these systems (lower panel of Figure \ref{fig:Fig_NSNS}), and therefore cannot aid them to merge. In NS-NS binaries that are expected to merge, though, a CBD could form relatively easily (Figure \ref{fig:allbeta}), potentially triggering a post-CEE merger of the NS with the exposed core of the giant star. Therefore, we conclude that in the case of NS-NS binaries, forming a CBD after CEE could only lower the event rates of NS-NS mergers.

\begin{figure}[h]
    \begin{center}
        \vspace{-0.2cm}
        \includegraphics[width=0.48\textwidth]{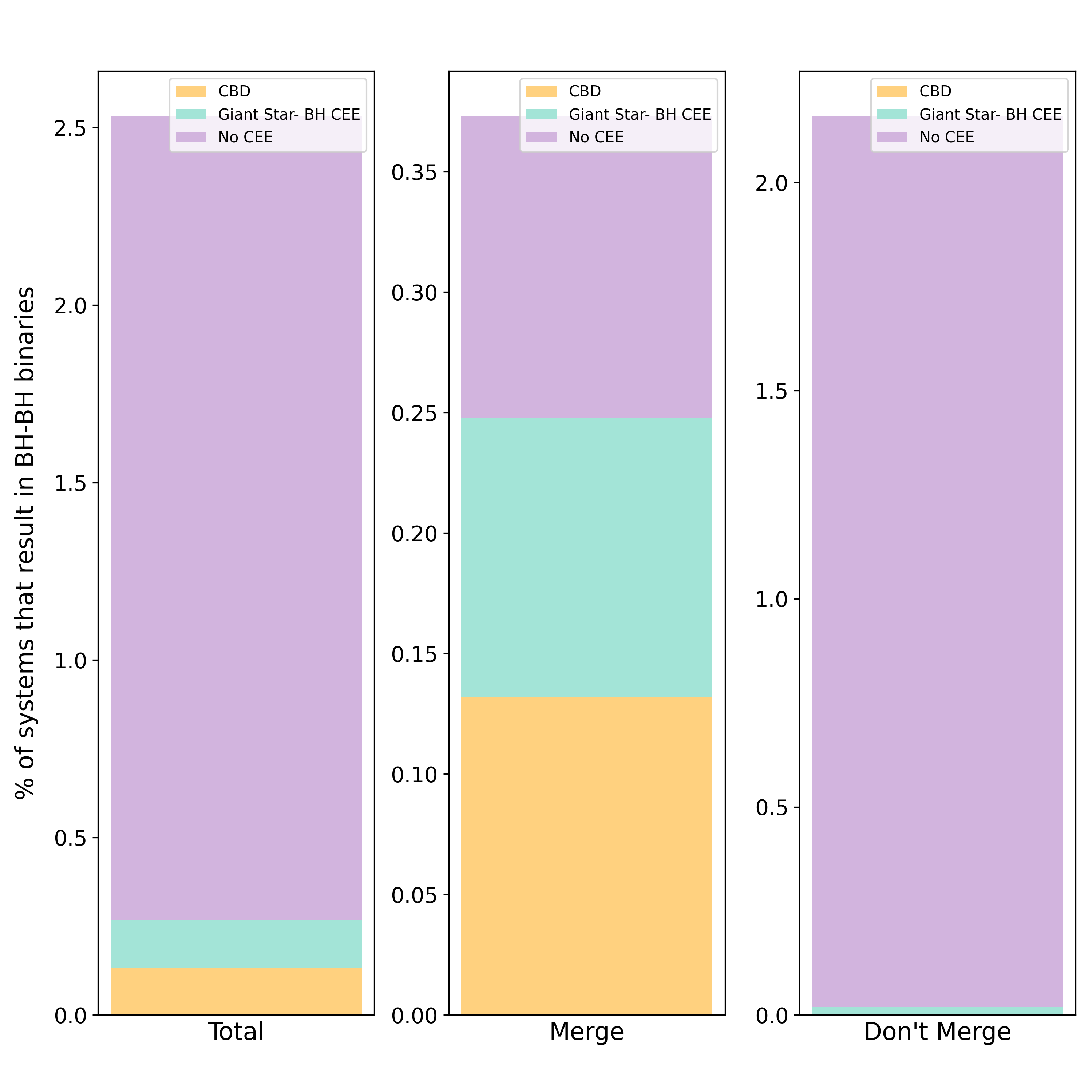}
        \caption{ Percentage of binary systems that evolve to BH-BH binaries from the overall number systems in our binary stellar population for a metallicity of $Z=0.01Z_{\rm \odot}$. Left panel: all BH-BH binaries that evolved through a CEE phase between the BH and the giant star (green columns), or that evolved via other channels (purple columns). Half of the BH-BH binaries form a post-CEE CBD (orange columns) for $\beta = 0.709$. Middle panel: same as left panel for BH-BH binaries that merge within Hubble time. Right panel: same as left panel for systems that are not expected to merge. } 
        \label{fig:BH_BH_Data_lower_Z}
    \end{center}
\end{figure}

We can see in Figure \ref{fig:BinaryData} that a large fraction of BH-BH binaries, and in particular most binaries that are not expected to merge (right panel), do not involve a CEE phase in their evolution, implying their overall formation rates will not be significantly affected by a potential post-CEE CBD. The lack of CBD formation in non-merging BH-BH binaries (lower panel of Figure \ref{fig:Fig_BHBH}) strengthens this conclusion and indicates that BH-BH merger rates are unlikely to increase when considering the formation of CBDs. For BH-BH binaries that merge, however, CEE is still the dominant formation channel according to our population synthesis study (see \citealt{Neijsseletal2019} and \citealt{GallegosGarciaetal2021} for another view). Considering a large fraction of these binaries are likely to form CBDs (Figure \ref{fig:allbeta}) that might shrink the orbit of the binary, in some cases BH-core mergers occur earlier in the evolution, reducing the number of BH-BH mergers. Figure \ref{fig:BH_BH_Data_lower_Z} shows that overall the same conclusion holds for a metallicity of $Z=0.01 Z_{\rm \odot}$ which is more typical for merging binary black holes, with the CEE being less prominent but with an higher overall number of binary black holes that evolve through CEE and result in DCO mergers (see section \ref{sec:Summary} for a discussion of the effects of metalicity on our results). 

Many NS-BH binaries go through a CEE of the BH and the giant (Figure \ref{fig:BinaryData}); in all these, the BH is the descendant of the primary star and the giant of the secondary star. Many of these end with CBD according to our criteria, implying that the CBD-induced early merger of the BH with the giant's core can reduce the number of NS-BH binaries. Contrary to NS-NS binaries, some post-CEE  non-merging NS-BH systems can form a CBD (lower panel of Figure \ref{fig:Fig_NSBH}). Losing angular momentum to the CBD will bring the stars in the surviving binary closer, resulting either in the merger of the core with the BH or forming a tighter NS-BH binary system that could merge in Hubble time. We note that there is a lower fraction of post-CEE non-merging NS-BH binaries compared to NS-BH binaries that are expected to merge. Therefore, it is more probable that the overall number of NS-BH mergers will decrease because of CBD-induced core-BH mergers. However due to contradictory effects, the merger rates will likely not change much.  

Overall, analyzing our results regarding the fraction of DCO post-CEE binaries and potential CBD formation in each class of systems, we conclude that if the orbital separation shrinks due to the binary's interaction with the CBD, the number of expected DCO mergers would be smaller compared to a case where CBD formation is neglected. The smallest effect is on the merger of NS-BH binaries. We note that if the orbital separation of the binary system grows due to angular momentum exchange with the CBD, this still results in a smaller number of DCO mergers. An increase in the orbital separation might drive the binary components far from each other, preventing a future merger. 

Another potential implication of the change in the orbital separation due to the interaction of the binary with the CBD would be changes in the predicted delay-time distributions of DCO mergers. If a sufficient number of NS-NS binaries contract, for instance, the steeper delay-time distribution (prompter mergers) might be able to contribute early enough NS-NS mergers to explain some of the heavy elements formation in the early Universe, as required by recent studies (e.g., \citealt{MaozNakar2024}; \citealt{BeniaminiPiran2024}). This is the subject of a future study

\section{Summary and discussion}
\label{sec:Summary}

In this study, we analyzed the population synthesis data of massive binaries from \cite{Grichener2023}, as described in Section \ref{sec:NumericalScheme}, to explore the formation of post-CEE CBDs in systems that evolve into DCO binaries. 

When an NS or BH engages in a CEE phase with a giant star, it spirals inside the envelope and ejects it. Some former envelope mass might become a close equatorial circumbinary gas, i.e., a CBD around the exposed giant's core and the compact object. Since this CBD formation process is challenging to simulate, we parametrized it with a toy model, as described in Section \ref{subsec:ToyModel}. We assume that the leftover envelope gas has a specific angular momentum that is a fraction $\beta$ of the total specific angular momentum at the onset of RLOF (equation \ref{eq:beta_updated}). If this specific angular momentum is larger than that of a Keplerian disk around the CBD's inner radius, then the left-over envelope forms a CBD (equation \ref{eq:jenvf}). In this exploratory study, we analyzed results for values of $\beta$ that lead half of NS-NS, BH-BH, and NS-BH binaries to form a CBD, as inferred from Figure \ref{fig:allbeta}.

CBDs will form least easily in NS-NS binary progenitors (Figure \ref{fig:Fig_NSNS} in Section \ref{sec:ResultsCBDs}). For DCO binaries that are not expected to merge within Hubble time, forming a post-CEE CBD is less likely due to the relatively large orbital separation at the end of CEE. In the case of NS-BH binaries, however, the larger mass ratio between the BH and the progenitor of the NS facilitates the formation of the CBD.
 
Our results affect the predicted formation and merger rates of DCO binaries. Angular momentum transfer between the CBD and the binary system dictates the final post-CEE orbital evolution (e.g., \citealt{TunaMetzger2023}; \citealt{Siweketal2023};  \citealt{Vallietal2024}; \citealt{Weietal2024}). We found that the number of NS-NS and BH-BH mergers will likely decrease independently of whether the binary widens or contracts due to interactions with the CBD. In cases where the orbital separation grows, the DCOs in the binary might drift so far apart that they may no longer be able to merge within Hubble time. If the CBD-binary interaction decreases the orbit, on the other hand, in some of the systems that were supposed to evolve into DCO mergers, the compact object and the core will merge shortly after the CEE. This likely results in luminous supernova-like transients when the core is composed of Helium or long gamma-ray bursts for more evolved cores. These transients are CEJSNe. We note that without accounting for CBD interaction, there is a negligible number of CEJSN that occur from a merger shortly after CEE. Therefore, a scenario where a CBD widens the orbit preventing a CEJSN and leading to a DCO mergers instead, will not increase the number of merging DCOs.

The possible contraction of the orbital separation might also result in shorter delay time distributions for DCO mergers, with potential implications for r-process nucleosynthesis, i.e., allowing NS-NS mergers to produce r-process elements at earlier stages of galactic evolution. Since we found most merging DCO binaries are expected to go through a CEE phase of a compact object and a giant star (middle panel of Figure \ref{fig:BinaryData}), we expect that CBD interactions would have a non-negligible effect in lowering the merger rates of DCO binaries. While quantifying the interaction between post-CEE CBDs and the binary systems they encompass to find the resultant orbital evolution and determining the lowering factor is a challenging task that is beyond the scope of this paper, we conclude it is important to keep in mind that CBD interaction might affect the rates of observables predicted by binary population synthesis simulations.  

Different choices of parameters and prescriptions in our population synthesis model can influence the results we obtain. Metallicity has been shown to have a non-negligible impact on the formation and merger rates of DCO binaries, with a greater influence on BH-BH formation that is expected to be dominant at lower metallicities (e.g. \citealt{Mapellietal2017}; \citealt{Giacobboetal2018}; \citealt{Neijsseletal2019}; \citealt{vanSonetal2024}; but see \citealt{Baveraetal2023}; \citealt{Lallemanetal2025L} for another view). We found that the formation and merger rates of BH-BH binaries grow significantly for $Z=0.01Z_{\rm \odot}$ with respect to solar metalicity (Fig. \ref{fig:BH_BH_Data_lower_Z}), with a much smaller effect on NS-NS and NS-BH binaries. A large fraction of the merging BH-BH systems ($\simeq 67 \%$) still evolve through CEE and most non-merging BH-BH binaries go through stable mass transfer and hence cannot form CBDs, implying our main conclusions from this study hold, and that we can predict fewer detected gravitational wave events with Ligo-Virgo-Kagra (LVK, e.g., \citealt{Abbottetal2018LVK}) compared to the expectations of population synthesis studies that neglect CBD formation.  Finding the exact reduction factor would require simulations with a metallicity-dependent star formation history, selection effects, and parametrization of the CBD-binary interaction. 

Mass loss through stellar winds significantly influences stellar evolution and the compact objects that massive stars form (e.g., \citealt{Smith2014MassLI}). Since \textsc{compas} does not account for wind accretion or wind interactions with the companion \citep{TeamCOMPAS2022},  we investigated the effect of conservative mass loss on CBD formation by running simulations accounting for accretion of stellar winds by adding to the accretor the mass lost by the donor. We found that wind accretion does not significantly impact our results. Other uncertainties related to binary evolution, such as the representation of CEE, tidal evolution, and natal kick prescriptions, might affect the results of our study in non-trivial ways and will be a subject of future studies. 

Despite these uncertainties, our study shows that accounting for post-CEE CBD formation in binary evolution is important for studying DCO formation and mergers. Incorporating prescriptions that allow studying CBD formation and interaction with the binary system in population synthesis could give insights into the expected observed rate of DCO binaries and transients that involve binaries with compact objects.  


\section*{ACKNOWLEDGMENTS}
\label{sec:acknowledgments}

We thank Ruggero Valli and Semih Tuna for the helpful discussions, and an anonymous referee for valuable suggestions and comments. AG acknowledges support from the Miriam and Aaron Gutwirth Fellowship, the Steward Observatory Fellowship in Theoretical and Computational Astrophysics, the IAU-Gruber Fellowship, and the CHE Fellowship. Simulations in this paper used the \textsc{compas} rapid binary population synthesis code (version 02.31.06), which is freely available at http://github.com/TeamCOMPAS/COMPAS. 

\section*{Data availability}
\label{sec:DataAvailability}
This manuscript made use of data generated by \cite{Grichener2023}, publicly available at Zenodo: \href{https://zenodo.org/records/11237180}{doi:10.5281}. In particular, we re-analyzed the population synthesis data from the folders $\rm alpha\_1$ $\rm Z\_solar$ and $\rm alpha\_1$ $\rm 0.01 Z\_solar$.

\bibliography{refs} 

\appendix

Let us consider a disk of surface density $\Sigma(r)$ for $r>R_{\rm CBD,in}$, which rotates with a Keplerian specific angular momentum around a binary of mass $M_{\rm bin}$. The disk's total mass and total angular momentum are 
\begin{equation}
\begin{split}
M = \int\Sigma(r) 2 \pi r dr,
\label{eq:MassDiskGeneral}
\end{split}
\end{equation}
and
\begin{equation}
\begin{split}
J = \int\Sigma(r) \sqrt{G M_{\rm bin} r} 2 \pi r dr,
\label{eq:JdiskGeneral}
\end{split}
\end{equation}
respectively. The mean specific angular momentum of the disk is then $j=J/M$.
For a surface density in the form of a power law $\Sigma(r)=\Sigma_{\rm 0} (r/r_{\rm 0})^{-m}$, direct integration gives
\begin{equation}
\begin{split}
M = \frac{2\pi \Sigma_{\rm 0} r_{\rm 0}}{m-2} \left( \frac{R_{\rm CBD,in}}{r_{\rm 0}} \right)^{-m+2},
\label{eq:MassDisk}
\end{split}
\end{equation}
and 
\begin{equation}
\begin{split}
J = \frac{2\pi \Sigma_{\rm 0} r_{\rm 0} \sqrt{G M_{\rm bin} r_{\rm 0}}}{m-2.5} \left( \frac{R_{\rm CBD,in}}{r_{\rm 0}} \right)^{-m+2.5},
\label{eq:Jdisk}
\end{split}
\end{equation}
where $m>2.5$ to keep the mass and angular momentum of the disk finite. 
The specific angular momentum of the disk is then
\begin{equation}
\begin{split}
j = \frac{m-2}{m-2.5} \sqrt{GM_{\rm bin}R_{\rm CBD,in}} = \zeta \sqrt{GM_{\rm bin}R_{\rm CBD,in}},
\label{eq:Jdisk}
\end{split}
\end{equation}
where $\zeta \equiv \frac{m-2}{m-2.5}$, implying that the specific angular momentum of the disk is a factor of $\zeta$ larger than the Keplerian specific angular momentum of a ring at $R_{\rm CBD,in}$.
\end{document}